\documentclass[twocolumn,prl,superscriptaddress]{revtex4}

\usepackage{graphicx}
\usepackage{dcolumn}
\usepackage{bm}
\usepackage{amstext}
\usepackage{color}

\begin{document}


\title{Separation of microscale chiral objects by shear flow}
\author{Marcos}
\affiliation{Department of Mechanical Engineering, MIT, Cambridge, MA 02139}
\author{Henry C. Fu}
\affiliation{Division of Engineering, Brown University, Providence, RI 02912}
\author{Thomas R. Powers}
\affiliation{Division of Engineering, Brown University, Providence, RI 02912}
\author{Roman Stocker}%
\affiliation{Department of Civil and Environmental Engineering, MIT, Cambridge, MA 02139}
\date{\today}

\begin{abstract}
We show that plane parabolic flow in a microfluidic channel causes nonmotile helically-shaped bacteria to drift perpendicular to the shear plane. Net drift results from the preferential alignment of helices with streamlines, with a direction that depends on the chirality of the helix and the sign of the shear rate. The drift is in good agreement with a model based on resistive force theory, and separation is efficient ($>80\%$) and fast ($<2$\,s). We estimate the effect of Brownian rotational diffusion on chiral separation and show how this method can be extended to separate chiral molecules.
\end{abstract}


\maketitle
Many biochemically active molecules are naturally chiral
and can only bind to target chiral molecules of a specific
handedness~\cite{Ahuja1997}. The other enantiomer (i.e. the molecule having opposite handedness) may be inactive or cause undesirable effects. Chemical synthesis of chiral molecules
usually produces a racemic mixture, with equal amounts of both
enantiomers, and their separation based on chirality is of
importance in fields ranging from agriculture to food and pharmaceutical industries. Currently favored approaches rely on gas, liquid or capillary electromigration chromatography~\cite{Scriba2008}, requiring costly chiral media. Thus, simpler, alternative approaches to chiral separation are desirable.

Several alternative proposals for chiral separation exploit hydrodynamic forces. Some of these, yet untested experimentally, rely on the presence of a surface~\cite{degennes99} or array of microvortices~\cite{Kosturetal2006}, and there has been successful chiral separation of cm-sized crystals in a rotating drum~\cite{Howardetal1976}. Other methods~\cite{KimRae1991, MakinoDoi2005} stem from the prediction that a chiral particle in a simple shear flow experiences a lateral drift~\cite{Brenner1964}. However, the feasibility of this approach has remained questionable, as measurements in Couette cells reported that the drift of mm-sized chiral objects~\cite{Makinoetal2008} and the forces on cm-sized ones~\cite{ChenChao2007} differ from predictions by two orders of magnitude~\cite{Makinoetal2008} or even in sign~\cite{ChenChao2007}.

Here we report that microscale chiral objects, three orders of magnitude smaller than previous studies~\cite{Makinoetal2008,
ChenChao2007}, experience a lateral drift in a microfluidic shear flow and the magnitude of the drift is in agreement with our theory. Previous work has demonstrated the ability of microfluidics to separate and sort colloids by size~\cite{dicarloetal2007}, spermatozoa by motility~\cite{choetal2003}, and microbes by their preference for dissolved chemicals~\cite{Stockeretal2008}.  Our method uses microchannels to sort particles by chirality. We show that an enantiomer drifts with direction determined by the local shear, demonstrate the feasibility of this method for chiral separation, and indicate how the high shear rates achievable in microchannels ($> 10^6$\,s$^{-1}$~\cite{Kangetal2005}) allow it to be extended to smaller scales ($<40$\,nm).

\begin{figure}
\includegraphics[width=8.5 cm]{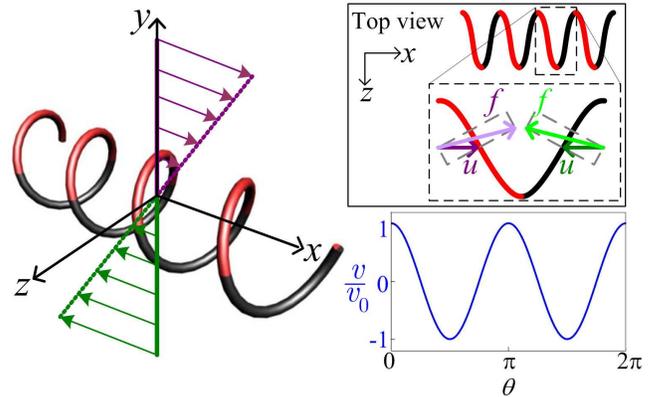}
\caption{(color) Schematic of a right-handed helix in simple shear flow. Red and black shadings show the top and bottom halves of the helix. Upper inset: the net force acting on one pitch of the helix is along $-\hat {\mathbf{z}}$. Lower inset: predicted normalized drift velocity $v/v_0$ vs. helix orientation $\theta$. $\theta$ is the angle between a helix in the $x$-$y$ plane and the flow, and $v_0$ is the drift velocity of a helix aligned with the flow ($\theta=0$).}\label{Figure1}
\end{figure}

\begin{figure*}
\includegraphics[width=17.6cm]{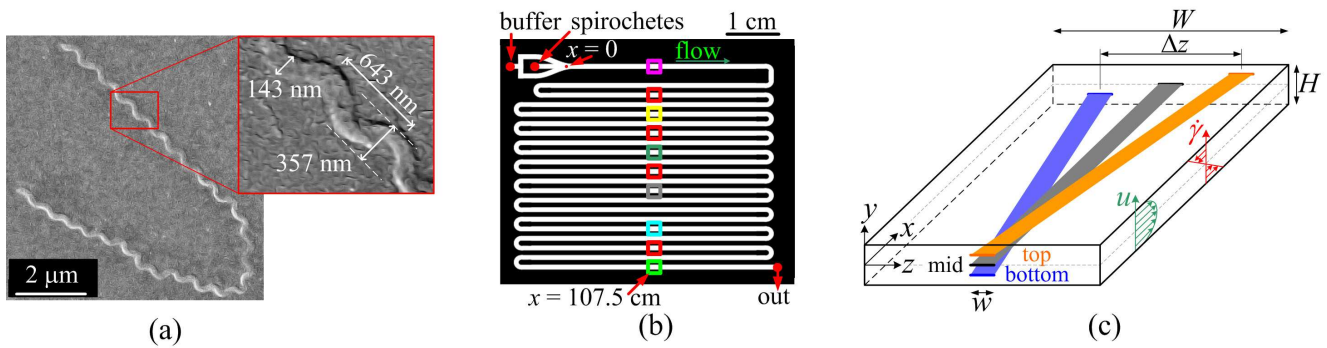}
\caption{(color) (a) Scanning Electron Micrograph (SEM) of \textit{L. biflexa}
 \textit{flaB} mutant, with typical dimensions (inset). The bent configuration
 is a result of SEM preparation and live organisms are
 nearly always straight. (b) Microchannel design with separate inlets
 for spirochetes and buffer. The color-coded squares refer to the
 locations of data collection (Fig. 3b). (c) Schematic of the
 separation process (for one enantiomer) in the microchannel ($W = 1\,\mathrm{ mm}$, $H =
 90\, \mu\mathrm{ m}$, $w \approx 100 \,\mu \mathrm{m}$). The lateral drift
 direction depends on the sign of the shear $\dot{\gamma}$, resulting in divergence
of top and bottom streams.
}\label{Figure2}
\end{figure*}

The origin of chirality-dependent drift at low Reynolds number can be simply understood for the case of a helix. In a shear flow, objects undergo periodic rotations known as Jeffery orbits~\cite{jeffery1922}: a sphere rotates with constant angular velocity, whereas for an elongated body, such as a helix, the velocity depends on orientation. The more elongated a body, the longer its residence time when aligned with streamlines. In addition to rotating in a Jeffery orbit, a helix drifts across streamlines. To see why, consider a right-handed helix aligned with a simple shear flow (Fig.~\ref{Figure1}), and decompose the velocity at a segment of the helix into components perpendicular and parallel to the segment. Drag on a thin rod in low Reynolds number flow is anisotropic, with a greater resistance when oriented perpendicular rather than parallel to the flow~\cite{Childress1981}. Since the flows at the top (red) and bottom (black) halves of the helix are in opposite directions, both halves have a drag component along
$-\hat{\mathbf{z}}$ (Fig.~\ref{Figure1}, top inset). Thus, a right-handed helix aligned with the shear flow drifts in the $-\hat{\mathbf{z}}$ direction. Reversing the chirality of the helix or the sign of the shear produces a drift along $+\hat{\mathbf{z}}$. Furthermore, the drift depends on the orientation of the helix. A right-handed helix aligned with the $y$ axis drifts along $+\hat{\mathbf{z}}$. In general, the drift velocity $v$ of a helix lying in the $x$-$y$ plane is a function of the angle $\theta$ between the helix and the flow (Fig.~\ref{Figure1}, bottom inset). Hence, the mean drift $\bar{v}$ depends on the distribution of the orientation of the helix relative to the flow. When the helix is preferentially aligned with streamlines, for example due to Jeffery orbits, it has a net drift. When all orientations are equally likely, for example due to strong Brownian rotational diffusion, the net drift is zero~\cite{MakinoDoi2005}. Here we use microscale objects to demonstrate the chirality-induced drift and determine the size of the smallest helices that can be separated using this principle.

We use a nonmotile, right-handed, helically-shaped bacterium, \textit{Leptospira biflexa} \textit{flaB} mutant~\cite{picardeau} (Fig.~2a), as the microscale chiral object. These spirochetes have an average length of $16$\,$\mu$m, thickness of $150\,\mathrm{nm}$, and diameter of $200\,\mathrm{nm}$ (Fig. \ref{Figure2}a and ~\cite{WolgemuthCharonGoldstein2006}). To expose spirochetes to a shear flow, we fabricated a 110 cm
long, serpentine-shaped microfluidic channel via soft
lithography~\cite{whitesidesetal2001} (Figs. \ref{Figure2}b,c). The channel has rectangular cross section of depth $H= 90 \,\mu\mathrm{m}$ and width $W = 1\, \mathrm{mm}$, resulting in a
parabolic flow, $u(y) = (3U/2) [1 - (2y/H)^{2}$], with mean speed $U = 3.09$ mm$\,$s$^{-1}$. Due to the high aspect ratio $W/H$, the velocity $u$
is uniform in $z$, except for a $150 \,\mu\mathrm{m}$ layer
adjacent to the two sidewalls~\cite{Doshietal1978}, inconsequential for these experiments. The shear rate $\dot {\gamma}(y)= du/dy = -12 yU/H^{2}$ varies linearly along
$y$ (between $\pm 206\,\mathrm{s}^{-1}$), with $\dot {\gamma }< 0$ for $y>0$,  and $\dot {\gamma }>0$  for $y<0$ (Fig.~\ref{Figure2}c). The microchannel is equipped with
two inlets, one for spirochetes and one for buffer
solution (EMJH liquid medium~\cite{picardeau}). Spirochetes were introduced
through a microinjector, forming a band of width $w \approx 100 \,\mu\mathrm{m}$ at the center of the channel, uniform over the depth, with
buffer streams on either side. As $\dot {\gamma }$ varies with $y$, one expects the drift to vary with vertical position, and spirochetes in the top and bottom halves of the channel to drift in opposite directions.

\begin{figure*}
\includegraphics[width=18cm]{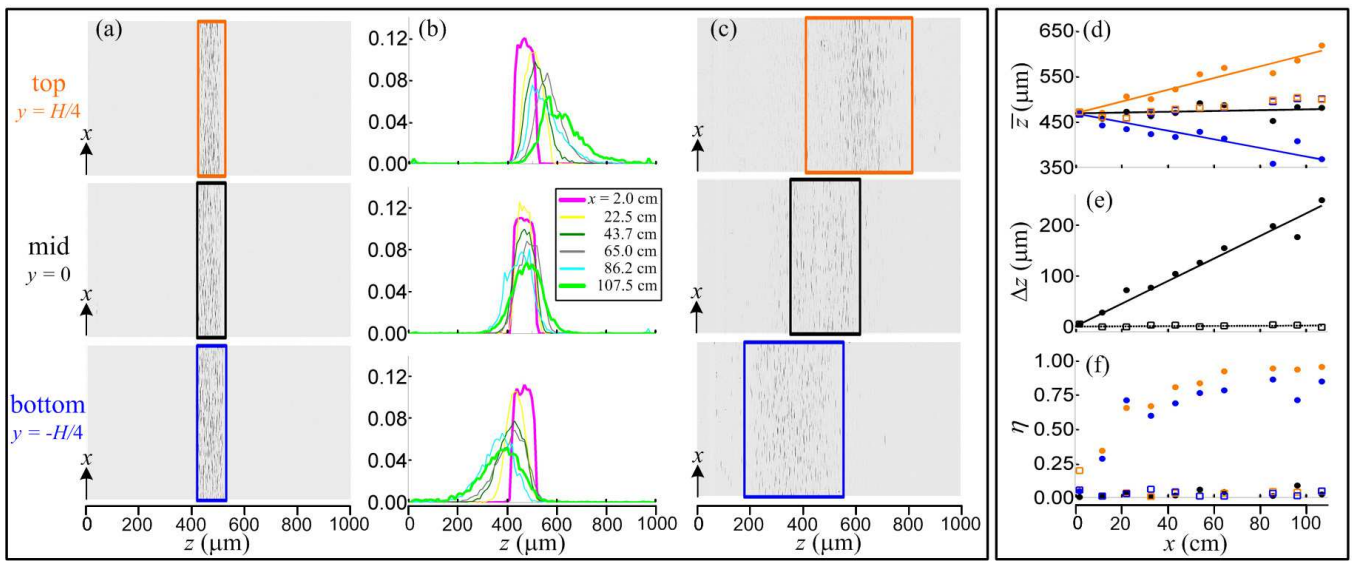}
\caption{(color) The observed spirochete distribution across the channel at (a) $x$ = 2
cm and (c) $x$ = 107.5 cm. Colors represent populations at different depths $y$: top (orange), mid (black) and bottom (blue) (see Fig. 2c). The mid point of each rectangle corresponds to the mean position ($\bar z$) of the distribution, while the half-width is two standard deviations ($2\sigma$). (b) Pdf of spirochete distribution at various distances $x$ along the
channel (see Fig. \ref{Figure2}b). The thick pink and green profiles
correspond to panels (a) and (c), respectively. (d,e,f) Quantification of
lateral drift and separation along $x$: (d) Mean position of the population at three depths. (e)
Divergence between the top and bottom populations. (f) Separation efficiency. In all panels, full circles refer to spirochete data, lines are linear fits, and empty squares are experiments with spherical beads. Over 10,000 spirochetes were imaged and located at each position $(x,y)$.
}\label{Figure3}
\end{figure*}

We imaged the spirochete population using phase contrast microscopy
with a 40$\times$ objective and a CCD camera. At ten locations along the channel (Fig.~\ref{Figure2}b), sets of fifty images were recorded at 1.5\,s intervals at three depths: $y=-H/4$ (``bottom''), 0 (``mid''), and $H$/4 (``top''). Experiments revealed a clear and reproducible drift, with direction determined by the sign of the local shear rate.  A superposition of
fifty images acquired at $x =107.5\, \mathrm{cm}$ shows drift
of comparable magnitude and opposite direction in the top and bottom halves, and no drift at mid-plane, where $\dot{\gamma}=0$ (Fig. \ref{Figure3}c and supplemental movie \cite{EPAPS}).
Using image analysis to obtain the $z$ position of individual spirochetes, we determined the probability density function (pdf) of the across-channel distribution of cells, their mean position $\bar z$, and the one-standard-deviation spread $\sigma$, for each imaging location $(x,y)$. The drift of the top and bottom populations increases monotonically with $x$ or, equivalently, with time. The apparent spread of the spirochete distribution with increasing $x$ (Fig. \ref{Figure3}b, c) is a result of the variation of drift with depth combined with the objective's finite depth of focus \cite{EPAPS}. Brownian motion was negligible ($\sigma \approx 6\,\mu\mathrm{m} $ over $x = 107.5 \,\mathrm{cm}$ \cite{EPAPS}).

Since drift ($\bar{z}$) increases linearly with distance $x$ along the
channel (Fig.~3d), the mean drift velocity is constant ($\bar{v} = 0.41 \, \mu $m s$^{-1}$), resulting
in a linearly increasing separation distance $\Delta z$
between the top and bottom populations (Fig.~\ref{Figure3}e).
Control experiments with neutrally buoyant $1\,\mu
\mathrm{m}$ spheres revealed no drift, irrespective of the vertical position in
the channel (Fig.~\ref{Figure3}d, empty squares), and thus no separation (Fig.~\ref{Figure3}e, empty squares), supporting the conclusion that drift is a result of chirality. We further verified that the 180-degree turns in the channel are inconsequential for these observations \cite{EPAPS}.

To quantify how the bottom and top streams diverge, we define a separation efficiency
$\eta$ = $\vert N_{L} - N_{R}\vert  /(N_{L} + N_{R})$, where
$N_{L}$ and $N_{R}$ are the numbers of spirochetes to the left and right
of the mean position $\bar z$ of the mid population ($y=0$),
and $\eta$ ranges from zero for no separation to one for perfect separation. We found $\eta$ to increase with $x$ for spirochetes, reaching a plateau at $\eta \approx 0.8$ after $x= 70 \,\mathrm{cm}$, while $\eta < 0.2$ for spheres (Fig. \ref{Figure3}f, empty squares). This separation was achieved in 227 s, and experiments with $U=600$ mm s$^{-1}$ ($\dot \gamma$ between $\pm$ 4 $\times$ 10$^4$ s$^{-1}$) further accelerated separation ($<$ 2 s). While we have separated a single enantiomer based on the sign of the shear, by symmetry our results also apply to separation of a racemic mixture in shear of a given sign.

\begin{figure}
\includegraphics[width=8.5cm]{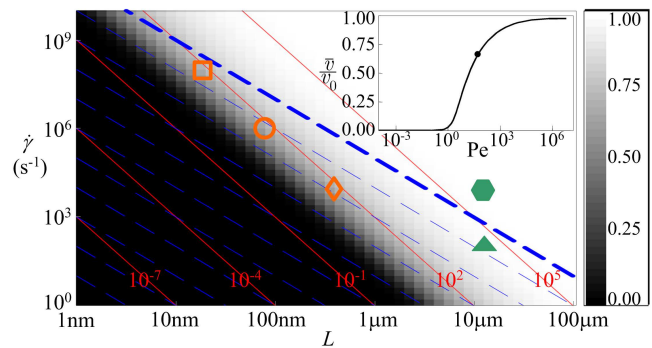}
\caption{(color online) Size limit for separation of isometric helices of length $L$ and equivalent aspect ratio $r$ = 70. Dashed contours show constant Re (spaced by a factor of 10) and the thick contour is Re = 0.1. Solid contours show constant Pe. The grayscale shows $\bar{v}/v_0$, where $v_0$ is the drift velocity of a helix aligned with the flow. The smallest helices which can be separated for a given shear rate, determined by $\bar{v}/v_0$ = 0.66 (Pe = 50, marked by circle on inset) and Re $<$ 0.1, have $L\approx$ 20 (square), 80 (circle), and 400 nm (diamond) for $\dot \gamma \approx 10^8, \, 10^6,\, \mathrm{and} \, 10^4 \, \mathrm{s}^{-1}$, respectively. Full symbols mark the parameter regimes of our experiments: Re = 0.03 (triangle) and Re = 5 (hexagon). Inset: $\bar{v}/v_0$ vs. Pe.
}\label{Figure4}
\end{figure}

To confirm that the observed lateral drift is due to the coupling of
shear and chirality, we use resistive force theory (RFT) ~\cite{Childress1981}
and model a spirochete as a 25-turn helix with dimensions shown in Fig.~\ref{Figure2}a. The drag force per unit length is $\mathbf{f} =
\zeta_{\parallel} \mathbf{v}^{r}_{\parallel} + \zeta_{\perp}
\mathbf{v}^{r}_{\perp}$, where $\zeta_{\parallel}$, $\zeta_{\perp}$ and
$\mathbf{v}^{r}_{\parallel}$, $\mathbf{v}^{r}_{\perp}$ are the resistive
coefficients and relative velocities parallel and perpendicular to
a given segment of the helix, respectively. The drag anisotropy
$\zeta_\perp/\zeta_\parallel$ depends on the thickness of the
spirochete body and  varies between 1.4 and 1.7~\cite{Childress1981}.
In Stokes flow, the linear and angular velocities of the
helix are determined by total force and total moment balance. This yields the drift velocity $v$ of a spirochete for an arbitrary orientation relative to the flow. Since we observe that the spirochetes lie in the $x$-$y$ plane to within 10 degrees, as expected on theoretical grounds~\cite{EPAPS}, we adopt a two-dimensional model with orientations limited to the $x$-$y$ plane. The mean drift velocity then depends on the probability distribution $c(\theta)$ of orientations of the helix relative to the flow. This distribution obeys $\partial_t c = D_R \, \partial^2_\theta c - \partial_\theta (\dot \theta c)$, where $D_{R} = 2.3 \times 10^{-3} \,\mathrm{s}^{-1}$ is the rotational diffusivity of an $L=$16\,$\mu$m long spirochete~\cite{EPAPS} and $\dot\theta$ is the rotation rate caused by the shear flow. We approximate $\dot\theta$ as the rotation rate of an ellipsoid with aspect ratio $r$, $\dot \theta = -\dot \gamma (\cos^2 \theta + r^2 \sin^2 \theta)/(1 + r^2)$~\cite{jeffery1922}, with $r=70$, as computed from the ratio of the rotation rates of the spirochetes at $\theta = 0$ and $\theta  =\pi/2$ obtained from RFT.
The steady-state solution ($\partial_t c = 0$) for $c(\theta)$ was obtained numerically and the mean drift velocity computed as $\bar{v} = \int_0^{2 \pi} v(\theta) c(\theta) \, \mathrm{d}\theta$. This resulted in $\bar{v} = 0.81$--$1.22 \, \mu$m~s$^{-1}$, which is of the same order of the measured value ($0.41 \, \mu $m s$^{-1}$). The residual discrepancy might be associated with the flexibility of the spirochetes and irregularities in their geometry.

For chirality-induced separation to take place, preferential alignment of helices with the flow is required. In our experiments, spirochetes were aligned hydrodynamically via Jeffery orbits. For smaller particles, rotational Brownian motion
is more effective at randomizing orientation. To determine the size of the smallest helical particles that can be separated in this manner, we find the steady distribution for helices with $r=70$ and length $L$. The mean drift velocity $\bar{v}$ is uniquely determined by the Peclet number $\mathrm{Pe} = \dot{\gamma}/D_R$ (Fig.~\ref{Figure4}, inset). For $\mathrm{Pe}\gg1$, alignment by shear dwarfs rotational diffusion and $\bar{v} \approx v_0$, where $v_0$ is the drift velocity of a helix aligned with the flow. Our experiments are in this limit, with $\mathrm{Pe} \approx 4 \times 10^{4}$ ($\dot{\gamma} \approx 103 \,\mathrm{s}^{-1}$ at $y=H/4$). In contrast, for $\mathrm{Pe}< 10$, diffusion destroys the alignment and
$\bar{v} \to 0$. Assuming isometrically scaled helices of length $L$, $D_{R}$ scales as $L^{-3}$, decreasing $\mathrm{Pe}$ for smaller helices. This can be partially counteracted by increasing $\dot{\gamma}$, but the particle Reynolds number $\mathrm{Re} = \dot{\gamma}L^{2}/\nu$~\cite{dicarloetal2007} must remain small ($\mathrm{Re}\ll1$) to operate in the Stokes regime. Hence, as $L$ decreases, $\dot{\gamma}$ can be increased as $L^{-2}$. The competition between $\mathrm{Pe}$ and $\mathrm{Re}$ determines the smallest helices that can be separated for a given shear rate, which we estimate as $L\approx$ 20, 80, and 400 nm for $\dot{\gamma} \approx 10^{8},\, 10^6$, and $10^4$ s$^{-1}$, respectively (Fig.~\ref{Figure4}, open symbols). Shear rates as high as $10^6$ s$^{-1}$ have been reported~\cite{Kangetal2005} in microchannels, and current technology allows  pressure heads of $2 \times 10^8$ Pa (Microfluidics Corp., Newton, MA), such that a 10 $\mu$m deep, 5 cm long channel could generate shear rates of 10$^7$ s$^{-1}$, sufficient to separate 40 nm particles.

While we have demonstrated the lateral drift of chiral
objects, one is ultimately interested in separating a mixture of
enantiomers. Because the top and bottom halves of the channel produce drift in
opposite directions, a racemic mixture would separate into four
quadrants, with opposite quadrants containing the same enantiomer. We are currently implementing three-dimensional microfabrication to separately collect particles
from the four quadrants. While we believe our technique represents a significant step forward in enabling reliable chiral separation by shear, it remains to be seen whether it can extend to the molecular scale ($<$ 10 nm). To this end, in addition to increasing shear rates via improved manufacturing techniques, it will be interesting to explore the feasibility of using alternative alignment mechanisms, such as external fields or boundary effects ~\cite{Makinoetal2008}, to counteract rotational diffusion. Alternatively, one
could operate at higher Reynolds number: preliminary experiments show successful
separation at Re = 5, but further research is required to elucidate the role of inertia on separation and any constraints on microchannel operation, such as Dean vortices and performance under high pressure. Finally, not all chiral particles are helical, so it would be useful to understand what geometries lead to the best separation.

\textbf{Acknowledgements} We thank M. Picardeau for providing spirochetes, H. Jang for assistance with SEM, and R.H.W. Lam, B.C. Kirkup, C. Schmidt and J.T. Locsei for discussions. Marcos was partially supported by NSF grants OCE-0526241 and OCE-0744641 CAREER to RS. TRP acknowledges partial support from NSF grant DMS-0615919.


\begin{thebibliography}{21}
\expandafter\ifx\csname natexlab\endcsname\relax\def\natexlab#1{#1}\fi
\expandafter\ifx\csname bibnamefont\endcsname\relax
  \def\bibnamefont#1{#1}\fi
\expandafter\ifx\csname bibfnamefont\endcsname\relax
  \def\bibfnamefont#1{#1}\fi
\expandafter\ifx\csname citenamefont\endcsname\relax
  \def\citenamefont#1{#1}\fi
\expandafter\ifx\csname url\endcsname\relax
  \def\url#1{\texttt{#1}}\fi
\expandafter\ifx\csname urlprefix\endcsname\relax\def\urlprefix{URL }\fi
\providecommand{\bibinfo}[2]{#2}
\providecommand{\eprint}[2][]{\url{#2}}

\bibitem[{\citenamefont{Ahuja}(1997)}]{Ahuja1997}
\bibinfo{author}{\bibfnamefont{S.}~\bibnamefont{Ahuja}},
  \emph{\bibinfo{title}{Chiral Separations: Applications and Technology}}
  (\bibinfo{publisher}{American Chemical Society}, \bibinfo{address}{Washington
  DC}, \bibinfo{year}{1997}).

\bibitem[{\citenamefont{Scriba}(2008)}]{Scriba2008}
\bibinfo{author}{\bibfnamefont{G.~K.~E.} \bibnamefont{Scriba}},
  \bibinfo{journal}{J. Sep. Sci.} \textbf{\bibinfo{volume}{{\bf 31}}},
  \bibinfo{pages}{1991} (\bibinfo{year}{2008}).

\bibitem[{\citenamefont{{de Gennes}}(1999)}]{degennes99}
\bibinfo{author}{\bibfnamefont{P.~G.} \bibnamefont{{de Gennes}}},
  \bibinfo{journal}{Europhys. Lett.} \textbf{\bibinfo{volume}{{\bf 46}}},
  \bibinfo{pages}{827} (\bibinfo{year}{1999}).

\bibitem[{\citenamefont{Kostur et~al.}(2006)\citenamefont{Kostur, Schindler,
  Talkner, and H\"{a}nggi}}]{Kosturetal2006}
\bibinfo{author}{\bibfnamefont{M.}~\bibnamefont{Kostur}},
  \bibinfo{author}{\bibfnamefont{M.}~\bibnamefont{Schindler}},
  \bibinfo{author}{\bibfnamefont{P.}~\bibnamefont{Talkner}}, \bibnamefont{and}
  \bibinfo{author}{\bibfnamefont{P.}~\bibnamefont{H\"{a}nggi}},
  \bibinfo{journal}{Phys. Rev. Lett.} \textbf{\bibinfo{volume}{{\bf 96}}},
  \bibinfo{pages}{014502} (\bibinfo{year}{2006}).

\bibitem[{\citenamefont{Howard et~al.}(1976)\citenamefont{Howard, Lightfoot,
  and Hirschfelder}}]{Howardetal1976}
\bibinfo{author}{\bibfnamefont{D.~W.} \bibnamefont{Howard}},
  \bibinfo{author}{\bibfnamefont{E.~N.} \bibnamefont{Lightfoot}},
  \bibnamefont{and} \bibinfo{author}{\bibfnamefont{J.~O.}
  \bibnamefont{Hirschfelder}}, \bibinfo{journal}{AICHE J.}
  \textbf{\bibinfo{volume}{{\bf 22}}}, \bibinfo{pages}{794}
  (\bibinfo{year}{1976}).

\bibitem[{\citenamefont{Kim and Rae}(1991)}]{KimRae1991}
\bibinfo{author}{\bibfnamefont{Y.-J.} \bibnamefont{Kim}} \bibnamefont{and}
  \bibinfo{author}{\bibfnamefont{W.~J.} \bibnamefont{Rae}},
  \bibinfo{journal}{Int. J. Multiphase Flow} \textbf{\bibinfo{volume}{{\bf
  17}}}, \bibinfo{pages}{717} (\bibinfo{year}{1991}).

\bibitem[{\citenamefont{Makino and Doi}(2005)}]{MakinoDoi2005}
\bibinfo{author}{\bibfnamefont{M.}~\bibnamefont{Makino}} \bibnamefont{and}
  \bibinfo{author}{\bibfnamefont{M.}~\bibnamefont{Doi}},
  \bibinfo{journal}{Phys. Fluids} \textbf{\bibinfo{volume}{{\bf 17}}},
  \bibinfo{pages}{103605} (\bibinfo{year}{2005}).

\bibitem[{\citenamefont{Brenner}(1964)}]{Brenner1964}
\bibinfo{author}{\bibfnamefont{H.}~\bibnamefont{Brenner}},
  \bibinfo{journal}{Chem. Eng. Sci.} \textbf{\bibinfo{volume}{{\bf 19}}},
  \bibinfo{pages}{631 } (\bibinfo{year}{1964}).

\bibitem[{\citenamefont{Makino et~al.}(2008)\citenamefont{Makino, Arai, and
  Doi}}]{Makinoetal2008}
\bibinfo{author}{\bibfnamefont{M.}~\bibnamefont{Makino}},
  \bibinfo{author}{\bibfnamefont{L.}~\bibnamefont{Arai}}, \bibnamefont{and}
  \bibinfo{author}{\bibfnamefont{M.}~\bibnamefont{Doi}}, \bibinfo{journal}{J.
  Phys. Soc. Japan} \textbf{\bibinfo{volume}{{\bf 77}}},
  \bibinfo{pages}{064404} (\bibinfo{year}{2008}).

\bibitem[{\citenamefont{Chen and Chao}(2007)}]{ChenChao2007}
\bibinfo{author}{\bibfnamefont{P.}~\bibnamefont{Chen}} \bibnamefont{and}
  \bibinfo{author}{\bibfnamefont{C.-H.} \bibnamefont{Chao}},
  \bibinfo{journal}{Phys. Fluids} \textbf{\bibinfo{volume}{{\bf 19}}},
  \bibinfo{pages}{017108} (\bibinfo{year}{2007}).

\bibitem[{\citenamefont{{Di Carlo} et~al.}(2007)\citenamefont{{Di Carlo},
  Irimia, Tompkins, and Toner}}]{dicarloetal2007}
\bibinfo{author}{\bibfnamefont{D.}~\bibnamefont{{Di Carlo}}},
  \bibinfo{author}{\bibfnamefont{D.}~\bibnamefont{Irimia}},
  \bibinfo{author}{\bibfnamefont{R.~G.} \bibnamefont{Tompkins}},
  \bibnamefont{and} \bibinfo{author}{\bibfnamefont{M.}~\bibnamefont{Toner}},
  \bibinfo{journal}{Proc. Nat. Acad. Sci. U.S.A.} \textbf{\bibinfo{volume}{{\bf
  104}}}, \bibinfo{pages}{18892} (\bibinfo{year}{2007}).

\bibitem[{\citenamefont{Cho et~al.}(2003)\citenamefont{Cho, Schuster, Zhu,
  Chang, Smith, and Takayama}}]{choetal2003}
\bibinfo{author}{\bibfnamefont{B.~S.} \bibnamefont{Cho}},
  \bibinfo{author}{\bibfnamefont{T.~G.} \bibnamefont{Schuster}},
  \bibinfo{author}{\bibfnamefont{X.}~\bibnamefont{Zhu}},
  \bibinfo{author}{\bibfnamefont{D.}~\bibnamefont{Chang}},
  \bibinfo{author}{\bibfnamefont{G.~D.} \bibnamefont{Smith}}, \bibnamefont{and}
  \bibinfo{author}{\bibfnamefont{S.}~\bibnamefont{Takayama}},
  \bibinfo{journal}{Anal. Chem.} \textbf{\bibinfo{volume}{{\bf 75}}},
  \bibinfo{pages}{1671} (\bibinfo{year}{2003}).

\bibitem[{\citenamefont{Stocker et~al.}(2008)\citenamefont{Stocker, Seymour,
  Samadani, Hunt, and Polz}}]{Stockeretal2008}
\bibinfo{author}{\bibfnamefont{R.}~\bibnamefont{Stocker}},
  \bibinfo{author}{\bibfnamefont{J.~R.} \bibnamefont{Seymour}},
  \bibinfo{author}{\bibfnamefont{A.}~\bibnamefont{Samadani}},
  \bibinfo{author}{\bibfnamefont{D.~E.} \bibnamefont{Hunt}}, \bibnamefont{and}
  \bibinfo{author}{\bibfnamefont{M.~F.} \bibnamefont{Polz}},
  \bibinfo{journal}{Proc. Natl. Acad. Sci. U.S.A.}
  \textbf{\bibinfo{volume}{{\bf 105}}}, \bibinfo{pages}{4209}
  (\bibinfo{year}{2008}).

\bibitem[{\citenamefont{Kang et~al.}(2005)\citenamefont{Kang, Lee, and
  Koelling}}]{Kangetal2005}
\bibinfo{author}{\bibfnamefont{K.}~\bibnamefont{Kang}},
  \bibinfo{author}{\bibfnamefont{L.~J.} \bibnamefont{Lee}}, \bibnamefont{and}
  \bibinfo{author}{\bibfnamefont{K.~W.} \bibnamefont{Koelling}},
  \bibinfo{journal}{Exp. Fluids.} \textbf{\bibinfo{volume}{{\bf 38}}},
  \bibinfo{pages}{222} (\bibinfo{year}{2005}).

\bibitem[{\citenamefont{Jeffery}(1922)}]{jeffery1922}
\bibinfo{author}{\bibfnamefont{G.~B.} \bibnamefont{Jeffery}},
  \bibinfo{journal}{Proc. Roy. Soc. A} \textbf{\bibinfo{volume}{{\bf 102}}},
  \bibinfo{pages}{161} (\bibinfo{year}{1922}).

\bibitem[{\citenamefont{Childress}(1981)}]{Childress1981}
\bibinfo{author}{\bibfnamefont{S.}~\bibnamefont{Childress}},
  \emph{\bibinfo{title}{Mechanics of swimming and flying}}
  (\bibinfo{publisher}{Cambridge University Press},
  \bibinfo{address}{Cambridge}, \bibinfo{year}{1981}).

\bibitem[{\citenamefont{Picardeau et~al.}(2001)\citenamefont{Picardeau, Brenot,
  and Girons}}]{picardeau}
\bibinfo{author}{\bibfnamefont{M.}~\bibnamefont{Picardeau}},
  \bibinfo{author}{\bibfnamefont{A.}~\bibnamefont{Brenot}}, \bibnamefont{and}
  \bibinfo{author}{\bibfnamefont{I.~S.} \bibnamefont{Girons}},
  \bibinfo{journal}{Mol. Microbiol.} \textbf{\bibinfo{volume}{{\bf 40}}},
  \bibinfo{pages}{189} (\bibinfo{year}{2001}).

\bibitem[{\citenamefont{Wolgemuth et~al.}(2006)\citenamefont{Wolgemuth, Charon,
  Goldstein, and Goldstein}}]{WolgemuthCharonGoldstein2006}
\bibinfo{author}{\bibfnamefont{C.~W.} \bibnamefont{Wolgemuth}},
  \bibinfo{author}{\bibfnamefont{N.~W.} \bibnamefont{Charon}},
  \bibinfo{author}{\bibfnamefont{S.~F.} \bibnamefont{Goldstein}},
  \bibnamefont{and} \bibinfo{author}{\bibfnamefont{R.~E.}
  \bibnamefont{Goldstein}}, \bibinfo{journal}{J. Mol. Microbiol. Biotechnol.}
  \textbf{\bibinfo{volume}{{\bf 11}}}, \bibinfo{pages}{221}
  (\bibinfo{year}{2006}).

\bibitem[{\citenamefont{Whitesides et~al.}(2001)\citenamefont{Whitesides,
  Ostuni, Takayama, Jiang, and Inger}}]{whitesidesetal2001}
\bibinfo{author}{\bibfnamefont{G.~M.} \bibnamefont{Whitesides}},
  \bibinfo{author}{\bibfnamefont{E.}~\bibnamefont{Ostuni}},
  \bibinfo{author}{\bibfnamefont{S.}~\bibnamefont{Takayama}},
  \bibinfo{author}{\bibfnamefont{X.}~\bibnamefont{Jiang}}, \bibnamefont{and}
  \bibinfo{author}{\bibfnamefont{D.~E.} \bibnamefont{Inger}},
  \bibinfo{journal}{Annu. Rev. Biomed. Eng.} \textbf{\bibinfo{volume}{{\bf
  3}}}, \bibinfo{pages}{335} (\bibinfo{year}{2001}).

\bibitem[{\citenamefont{Doshi et~al.}(1978)\citenamefont{Doshi, Daiya, and
  Gill}}]{Doshietal1978}
\bibinfo{author}{\bibfnamefont{M.~R.} \bibnamefont{Doshi}},
  \bibinfo{author}{\bibfnamefont{P.~M.} \bibnamefont{Daiya}}, \bibnamefont{and}
  \bibinfo{author}{\bibfnamefont{W.~N.} \bibnamefont{Gill}},
  \bibinfo{journal}{Chem. Eng. Sci.} \textbf{\bibinfo{volume}{{\bf 33}}},
  \bibinfo{pages}{795} (\bibinfo{year}{1978}).

\bibitem[{EPA()}]{EPAPS}
\bibinfo{note}{See EPAPS Document No. [number will be inserted by publisher]
  for supplemental material and movie. For more information on EPAPS, see
  http://www.aip.org/pubservs/epaps.html.}

\end{thebibliography}
\end{document}